\documentclass[a4paper]{article}
\usepackage{amssymb}
 \usepackage{graphicx}
 \newcommand{\singlefig}{.75\textwidth}
 
\title
{\bf Charge transport in poly(dG)-poly(dC) and poly(dA)-poly(dT)
DNA polymers}
\author{{\bf D. Hennig$^{1}$, E.B. Starikov$^{2}$, J.F.R. Archilla$^{3}$ and
F. Palmero$^{3}$}
\\
$^1$ Freie Universit\"{a}t Berlin, Fachbereich Physik, Institut
f\"{u}r Theoretische Physik
\\
Arnimallee 14, 14195 Berlin, Germany
\\
$^2$ Karolinska Institute,
Center for Structural Biochemistry NOVUM\\
S\,-\,14157 Huddinge, Sweden\
\\
$^3$ Group of Nonlinear Physics, Departamento de F\'{i}sica
Aplicada I\\ ETSI Inform\'{a}tica,  Avda Reina Mercedes, s/n.
41012\,-\,Sevilla, Spain\\}
\date{August 1, 2003}
\begin{document}
 \maketitle

\begin{abstract}
\noindent We investigate the charge transport in synthetic DNA
polymers built up from single types of base pairs. In the context
of a polaron-like model, for which an electronic tight-binding
system and bond vibrations of the double helix are coupled, we
present estimates for the electron-vibration coupling strengths
utilizing a quantum-chemical procedure. Subsequent studies
concerning the mobility of polaron solutions, representing the
state of a localized charge in unison with its associated helix
deformation, show that the system for poly(dG)-poly(dC) and
poly(dA)-poly(dT) DNA polymers, respectively  possess
quantitatively distinct transport properties. While the former
supports unidirectionally moving electron breathers attributed to
highly efficient long-range conductivity the breather mobility in
the latter case is comparatively restrained inhibiting charge
transport. Our results are in agreement with recent experimental
results demonstrating that poly(dG)-poly(dC) DNA molecules
 acts as a semiconducting nanowire and exhibits
better conductance than poly(dA)-poly(dT) ones.
\end{abstract}


\section{Introduction}

Particularly  with view to possible applications in molecular
electronics based on biomaterials electronic transport (ET)
through DNA has recently become of intensified interest
\cite{Ratner},\cite{Kasumov}. Although the debate whether DNA
constitutes a conductor is still ongoing, there exist already
strong experimental evidence that DNA forms an effectively
one-dimensional molecular wire \cite{Fink}. Among several
theoretical attempts to describe the charge transport mechanism in
DNA the polaron approach has turned out lately to be a promising
candidate for modeling constructive interplay between the charge
carrying system and vibrational degrees of freedom of the DNA
conspiring to establish coherent ET \cite{Bruinsma}-\cite{Basko}.
Recent experiments are in support of the polaron mechanism for ET
in DNA polymers \cite{Kawai}. The present study deals with the
theoretical description of ET in synthetically produced  DNA
polymers consisting of a single type of base pairs, i.e. either
poly(dG)-poly(dC) or poly(dA)-poly(dT) DNA polymers \cite{Porath}.
Utilizing a nonlinear approach based on the concept of breather
and polaron solutions we explore whether conductivity depends on
the type of the DNA polymer which might be of interest for the
design of synthetic molecular wires. Moreover, we ameliorate
preceding studies of ET in DNA \cite{PhysicaD},\cite{control} in
the sense that, instead of adjusting the coupling parameters, we
use now credible estimates for them derived with the help of
quantum-chemical methods.

\section{Model for polaron-like charge transport in DNA}
Our model of charge transport in DNA is based on the finding that
the charge migration process is dominantly influenced by the
transverse vibrations of the bases relative to each other in
radial direction within a base pair plane \cite{Stryer}. In fact,
the impact of other vibrational degrees of freedom (e.g. twist
motions, helical pitch changes and longitudinal acoustic phonons
along the strands which are significantly restrained by the
backbone rigidity) can be expected to be negligible with respect
to ET in DNA. Hence, the motion can be viewed as confined to the
base pair planes \cite{Jessica}.

The Hamiltonian for the ET along a strand in DNA  comprises two
parts
\begin{equation}
H=H_{el}+H_{vib}\,,
\end{equation}
with $H_{el}$ is  the part which describes  the ET over the base
pairs and $H_{vib}$ represent  the  dynamics of radial vibrations
of the base pairs . The electronic part is given by a
tight-binding system
\begin{equation}
H_{el}=\sum_{n} \,E_{n}\,|c_{n}|^2 -V_{n\,n-1}\,\left(\,c_{n}^{*}
c_{n-1}+c_n c_{n-1}^{*}\,\right) \,.\label{eq:Hel}
\end{equation}
The  index  $n$ denotes the site of the $n-$th base on a strand
and $|c_n|^2$ determines the probability to find the electron
(charge) residing at this site. $E_{n}$ is the local electronic
energy and $V_{nn-1}$ is the transfer matrix element mediating the
transport of the electron along the stacked base pairs. We make
the usual assumption that ET takes place only along the base pair
sequence on a strand excluding inter-strand ET.

The vibronic part of the Hamiltonian $H_{vib}$ models dynamical
changes of the radial equilibrium positions of the bases.
Supposing that these radial vibrations can be treated classically
and harmonically we represent $H_{vib}$ as
\begin{equation}
H_{vib}=
\frac{1}{2}\,\sum_{n}\,\left[\,\frac{1}{M}\left(p^{\,r}_{n}\right)^{\,2}
\,+\,M\,\Omega_r^2\,r_n^2\,\right]\,.
\end{equation}
The radial coordinates $r_{n}$ quantify the radial displacements
of the base units from their equilibrium positions along the line
bridging two bases of a base pair within the base pair plane. $M$
denotes the reduced mass and $\Omega_r$ is the harmonic frequency.
Due to the constraint enforced by the sugar-phosphate backbone
radial displacements lead not only to changes of the equilibrium
lengths of the hydrogen bonds (bridging two bases of a base pair)
but are also connected with a change of the three-dimensional
distance between two consecutive bases on a strand entailing
deformations of the corresponding covalent bonds. In an expansion
up to first order around the equilibrium positions \cite{Peyrard}
this stacking-distance is determined by
\begin{equation}
d_{n\,n-1}=\frac{R_0}{l_0}\,(\,1-\cos
\theta_0\,)\,(r_n+r_{n-1})\,.\label{eq:distance}
\end{equation}
The geometrical parameters $R_0$, $\theta_0$ and $l_0$ determine
the equilibrium value of the radius, the twist angle between two
adjacent base pairs, and the equilibrium length of the covalent
bond linking two consecutive bases along a strand, respectively.
The latter is given by
\begin{equation}
l_0=\sqrt{a^2+4R_0^2\,\sin^2(\theta_0/2)}\,,\label{eq:l0}
\end{equation}
with $a$ being the distance between neighboring base pairs
measured along the orientation of the helix axis. A sketch of
the structure of the DNA model and the designation of the
geometrical parameters $R_0$, $l_0$, $\theta_0$, $r_n$ and
$d_{n\,n-1}$ is presented in Figure $1$.

\begin{figure}
  \begin{center}
 \includegraphics[width=\singlefig]{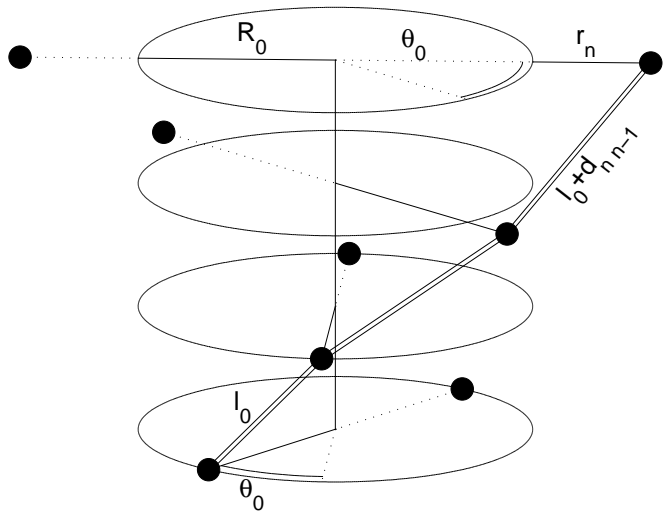}
  \end{center}
  \caption{Sketch of the structure of the DNA model. The bases are represented by bullits
and the geometrical parameters $R_0$, $l_0$, $\theta_0$, $r_n$ and
$d_{n\,n-1}$ are indicated.}
  \label{model}
\end{figure}

With respect to the interaction between the electronic and the
vibrational degrees of freedom variable,  $c_n$ and $r_n$,
respectively, it is assumed that the electronic parameters $E_n$
and $V_{n\,n-1}$ are modified by  displacements of the bases
within the base pairs. As a quantum-chemical computation of the
geometry dependence of the electronic parameters $E_n$ and
$V_{n\,n-1}$ reveals their most significant modulation originates
from radial distortions of the helix  which are related with
hydrogen and covalent bond deformations, respectively. However,
the influence of small angle deformations on the values of the
electronic parameters can be discarded (see also further below).

The modulation of the on-site electronic energy $E_{0}$ by the
radial vibrations of the base pairs is expressed as
\begin{equation}
E_{n}=E_{0}+k\,r_{n}\,.
\end{equation}
On the other hand the actual charge occupation has its impact on
the local radial distortion of the helix.  Furthermore, the
transfer matrix elements $V_{n\,n-1}$ are assumed to depend on the
three-dimensional stacking-distance between two consecutive bases
on a strand as follows
\begin{equation}
V_{n\,n-1}=V_0\,(1-\alpha\,d_{n\,n-1})\,.
\end{equation}
The quantity $\alpha$ regulates how strong $V_{n\,n-1}$ is
influenced by the distance.

As typical parameters for DNA molecules one finds
\cite{Stryer},\cite{Peyrard}: $a=3.4\AA$, $R_0 \thickapprox
10\AA$, $\theta_0=36^\circ$, $E_{0}=0.1\,eV$, $\Omega_{r}=6.252\,
\times10^{12}\,s^{-1}$, $V_0\simeq 0.1\,eV$ and
$M=4.982\times10^{-25}kg$.

After scaling the time as $t\rightarrow \Omega_{r}\,t$ one passes
to the dimensionless quantities:
\begin{eqnarray}
\tilde{r}_{n}&=&\sqrt{\frac{M \Omega_r^{2}}{V_{0}}}\, r_{n}\,,
\qquad \tilde{k}=\frac{k}{\sqrt{M\Omega_r^{2}V_0}}\,,\qquad
\tilde{E_0}=\frac{E_0}{V_0}\,,\\
\tilde{\alpha}&=&\sqrt{\frac{V_0}{M\,\Omega_r^2}}\,\alpha\,,
\,\,\,\, \tilde{R}_{0}=\sqrt{\frac{M\,\Omega_r^2}{V_0}}\,R_{0} \,,
\end{eqnarray}
and for the sake of convenience the tildes are omitted in the
following. The values of the scaled parameters  are obtained as
 $R_0=34.862$ and $l_0=24.590$.

The set of coupled equations of motion read as
\begin{eqnarray}
i\,\tau\dot{c}_{n}&=&(E_{0}\,+k\,r_n)\,c_n\,\nonumber\\
&-&(1-\alpha\,d_{n+1,n})\,c_{n+1} -(1-\alpha\,d_{n\,n-1})\,c_{n-1}
\label{eq:dotc}\\
\ddot{r}_{n}&=&-r_n-k\,|c_n|^2\,-\,\frac{R_0}{l_0}\,(1-\cos
\theta_0)\,\nonumber\\ &\times&
\,\alpha\left(\,[c_{n+1}^*c_{n}+c_{n+1}c_{n}^*]+
[c_{n}^*c_{n-1}+c_{n}c_{n-1}^*]\,\right)\label{eq:dotr}
\end{eqnarray}
and the ratio $\tau=\hbar\,\Omega_{r}/V_0=0.0411$ determines the
time scale separation between the slow electron motion and the
fast bond vibrations.  (Notice that any $E_{0}c_n$ term on the
r.h.s. of Eq. (\ref{eq:dotc}) can be eliminated by a gauge
transformation $c_{n}\rightarrow \exp(-iE_{0}t/\tau)c_{n}$.)

Contrary to previous studies we use for our computations credible
values for the electron-mode coupling strengths $k$ and $\alpha$
as a result of our a quantum-chemical computational procedure.

In this work, we perform quantum-chemical calculations on
symmetrical homodimers consisting of two nucleoside Watson-Crick
base pairs (adenosine-thymidine (AT) and guanosine-cytidine (GC)
base pair steps (BPS)) stacked over each other to mimic the
conventional A-DNA and B-DNA conformations (for a schematic view
see \cite{Saenger}).
Taking into account DNA backbone at least in form of intact sugar
moieties, instead of substituting it by protons or methyl groups,
is necessary for the consistency of the calculations
\cite{Brunaud} and to correctly describe charge transfer through
DNA duplexes \cite{Cuniberti}. We use semiempirical
all-valence-electron PM3 Hamiltonian \cite{Stewart} within the
MOPAC7 version of CI (configuration interaction) approximation, as
described in detail in \cite{Starikov}. We chose here PM3-CI
method, but not an ab initio one, since

\noindent a) The molecular fragments involved are very large;

\noindent b) Similar ab initio calculations even for smaller
segments experience difficulties with SCF convergence
\cite{Fortunelli};

\noindent c) The work \cite{Brunaud} used similar semiempirical
Hamiltonian (AM1) for analogous molecular fragments;

\noindent d) PM3-CI approximation is good at describing excited
states of nucleoside base pair steps \cite{Starikov};

\noindent e) CI approximation is indispensable when charged states
of nucleic acid bases are considered using semiempirical quantum
chemistry (cf., e.g., \cite{Chen} and references therein).

Here the effects of small, but non-negligible (up to 0.1 Å),
radial stretching and compressing of Watson-Crick hydrogen bonds
(WC H-bonds) on the MO energies of the BPS under study have been
estimated. Specifically, we perturbed in the above sense the
equilibrium lengths of the WC H-bonds in only one of the base
pairs in all the BPS involved and monitored the resulting changes
in the energies of the highest occupied molecular orbital and of
the occupied molecular orbital next to the highest one (HOMO and
HOMO-1, respectively), as compared with the equilibrium values of
these energies. According to the Koopmans theorem well known in
quantum chemistry (cf., e.g., \cite{Fortunelli} and references
therein), the HOMO energy is approximately equal to molecular
ionisation potential and in the tight-binding approximation could
be viewed as the site energy, whereas the difference between the
HOMO and HOMO-1 energies is approximately twice the hopping
integral. With this in mind, we were able to estimate $k$ and
$\alpha$ parameters of our model by calculating linear regression
of the corresponding site energy and hopping integral changes,
respectively, onto WC H-bond distance perturbations. We were able
to arrive at a very tight linear correlation between the former
and the latter ones. Interestingly, we failed to reveal any
correlation between the tight-binding Hamiltonian parameter
changes and BPS twist angle/helical pitch perturbations ('helical
pitch' is the distance $a$ in Eq. (\ref{eq:l0})). The
perturbations in the two latter parameters were also relatively
small (up to 10 degrees and 0.1 Å, respectively). For the coupling
parameters of the poly(dA)-poly(dT) DNA polymer we obtain:
$k=0.0778917\,eV/\,\AA$ and $\alpha=0.053835\,\AA^{-1}$.  The
corresponding values for the poly(dG)-poly(dC) DNA polymer are
determined as $k=-0.090325\,eV/\,\AA$ and
$\alpha=0.383333\,\AA^{-1}$.

That we are equipped with the quantum-chemical estimates of the
coupling parameters $\alpha$ and $k$ is a definite step forward
and distinguishes the present study
 from previous ones \cite{PhysicaD},\cite{control}. We emphasize that the quantum-chemical
estimates for the coupling parameters differ significantly from
the ones used for the model study in
\cite{PhysicaD},\cite{control} where these parameters, in lack of
reliable values for them, were treated as adjustable. As a
consequence there is a difference between the ET scenario
described in \cite{PhysicaD},\cite{control} and the one we are
going to illustrate in the following.

\section{Stationary localized  electron-vibron states}

Caused by the nonlinear interplay between the electronic and the
vibrational degrees of freedom of the helix the formation of
polaronic electron-vibration compounds is possible
\cite{Bruinsma}-\cite{Basko},\cite{PhysicaD},\cite{control}. We
construct such localized stationary solutions of the coupled
system (\ref{eq:dotc}),(\ref{eq:dotr})
 with the help of the nonlinear map approach explained
 in detail in \cite{Kalosakas}. In Figure $2$ we depict the profiles of
 the (standing) polaron states for the poly(dA)-poly(dT) and
 the poly(dG)-poly(dC) DNA polymer, respectively.  In both cases the
 polarons are of fairly large extension (width).
 Regardless of the DNA polymer type the electronic wave function is
 localized  at lattice site and the envelope of the
 amplitudes decays monotonically and
exponentially with growing distance from this central site (base
pair).
 However, the
 electronic wave function of the poly(dG)-poly(dC) DNA polymer is
 stronger localized than the one of its poly(dA)-poly(dT) counterpart.
\begin{figure}
  \begin{center}
 \includegraphics[angle=-90,width=0.49\textwidth]{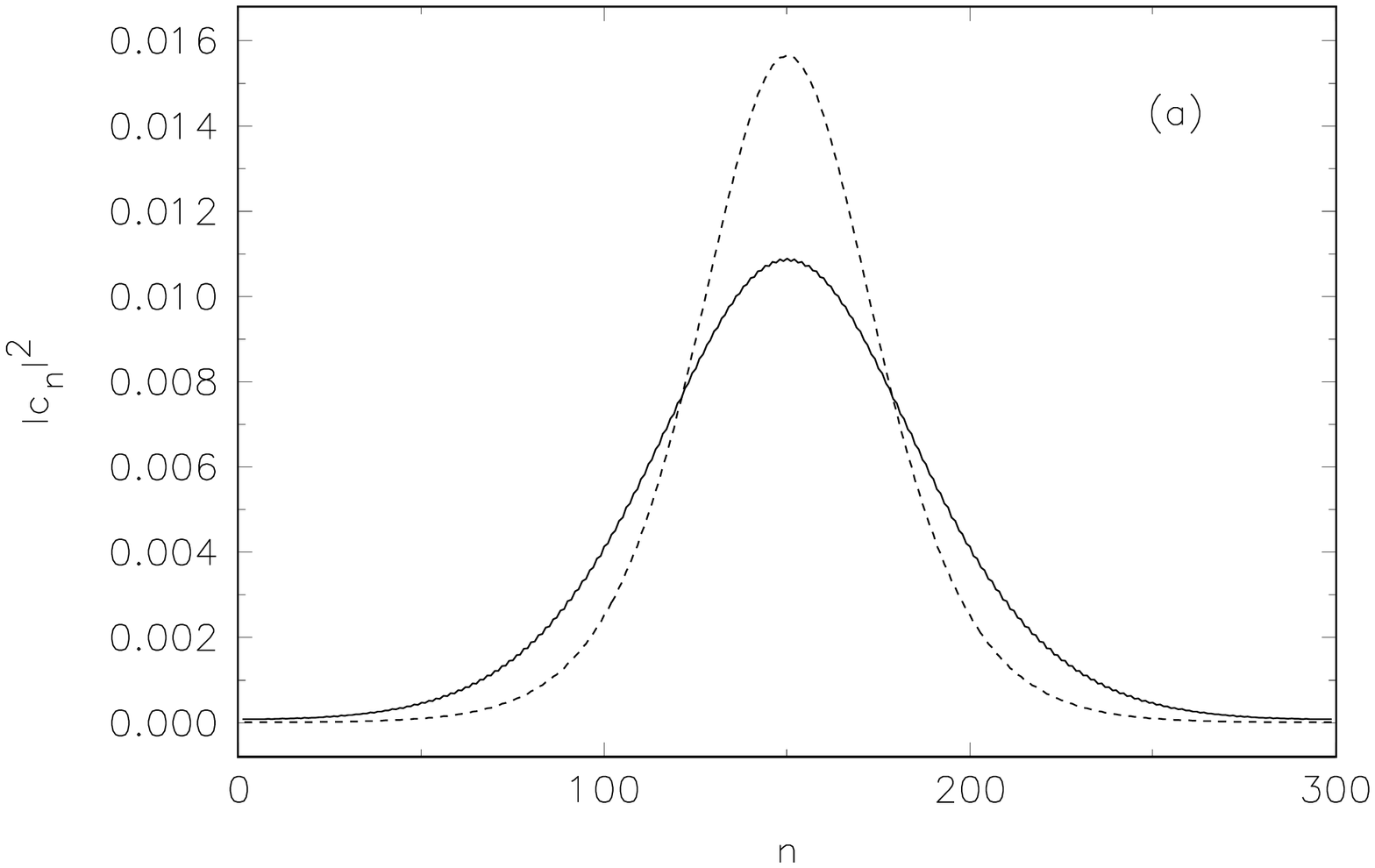}
 \includegraphics[angle=-90,width=0.49\textwidth]{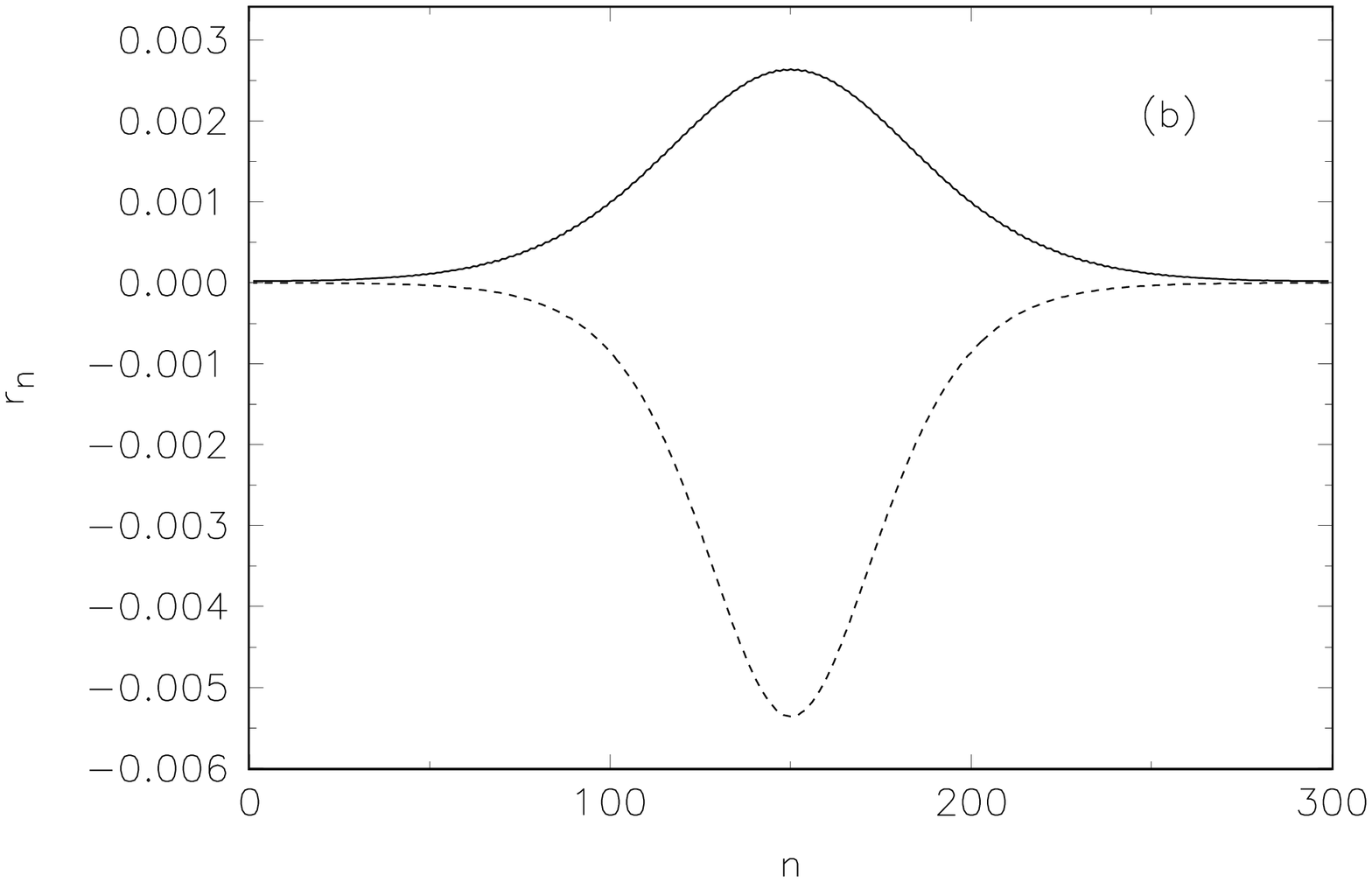}
  \end{center}
  \caption{The spatial pattern of the polaronic
electron-vibration compound. (a) The electronic part. Full
(dashed) line: poly(dA)-poly(dT) (poly(dG)-poly(dC)) DNA polymer.
(b) Unitless radial deformation pattern. Assignment of line types
as in (a). }
\end{figure}

 Accordingly, the attributed radial displacement patterns
  are exponentially localized at the central lattice site.
Concerning the resulting static radial helix deformations we find
that there is a drastic difference between the poly(dA)-poly(dT)
and the poly(dG)-poly(dC) DNA polymers. In the former case the
overall non-positive radial amplitudes imply that the H-bridges
experience contractions. In contrast, in the latter case the
H-bridges get stretched. Nevertheless, these deformations are
rather weak, i.e. on the order of $1.5\,10^{-3}\,\AA$.

\section{Charge transport}

We study now the ET achieved by moving polarons. In fact, since
the constructed polaron solutions are of fairly large extension
(the half-width involves $\lesssim 100$ lattice sites) we can
expect them to be mobile. In order to activate polaron motion we
used the discrete gradient method \cite{GTS} to obtain suitable
initial perturbations of the momentum coordinates $p_n^{\,r}$
which initiate coherent motion of the polaron compound.

We consider first the case of the poly(dG)-poly(dC) DNA polymer.
The propagation features are illustrated in Figure $3$ where the
spatio-temporal evolution of the electronic and the vibrational
radial breather are shown. For a DNA lattice consisting of $300$
sites (base pairs) the set of coupled equations
(\ref{eq:dotc}),(\ref{eq:dotr}) was integrated using a
fourth-order Runge-Kutta method and periodic boundary conditions
were imposed. Maintenance of the norm conservation
$\sum_{n}\,|c_n(t)|^2=1$ served to assure accurate computations.
\begin{figure}
  \begin{center}
 \includegraphics[angle=-90,width=0.9\textwidth]{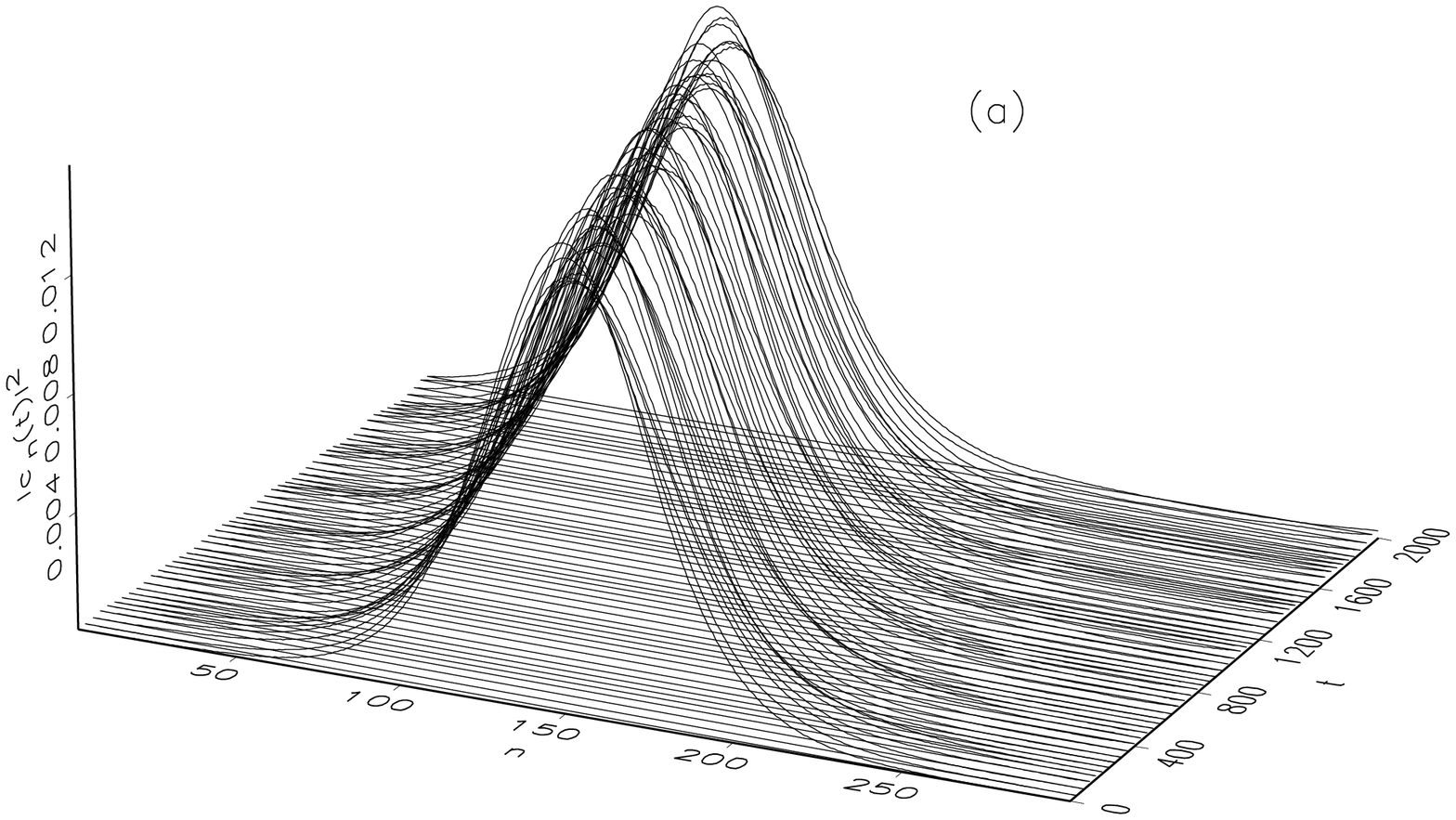}\\
 \includegraphics[angle=-90,width=0.9\textwidth]{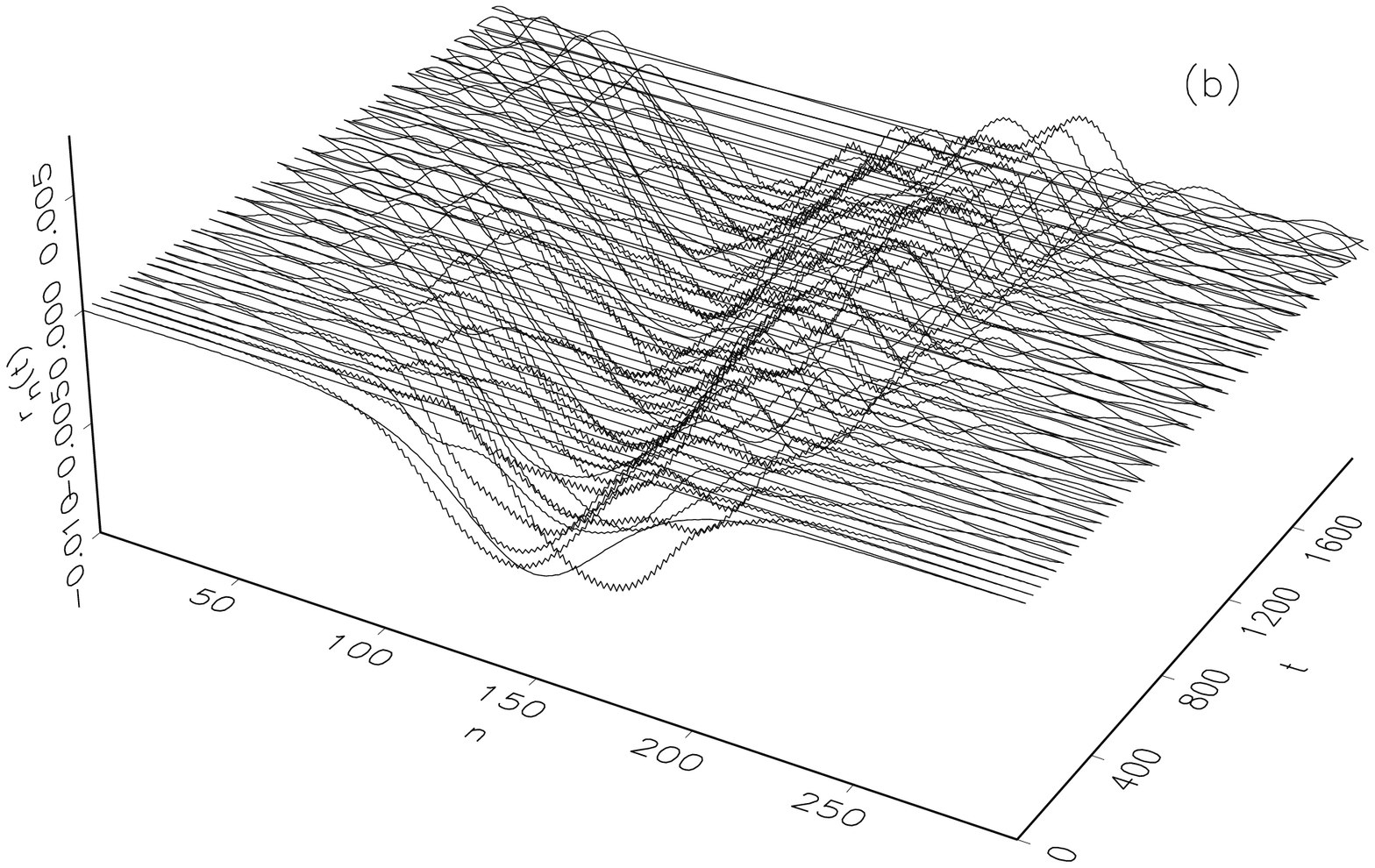}
  \end{center}
  \caption{Breather motion along the DNA for the
poly(dG)-poly(dC) DNA polymer.(a) The electronic breather. (b) The
vibrational breather.  The radial deformations $r_n(t)$ are given
in dimensionless units. }
\end{figure}

The electronic component sets off to move directionally along the
lattice in unison with the vibrational amplitude pattern. However,
there remains a small amplitude vibrational breather at the
initial position. In both cases the localized structures are
practically preserved apart from the arising amplitude breathing
indicating a periodic energy exchange between the electronic and
vibrational degrees of freedom. Hence, long-range ET is
achievable. With the initial injection of kinetic energy the
(vibrational) system possesses now increased energy content so
that we observe vibrational breathers with amplitudes being larger
than those of their static equivalents (cf. Figures $2\,(b)$ and
$3\,(b)$).

As the poly(dA)-poly(dT) DNA polymer is concerned we found that it
does not exhibit such good conductivity as its poly(dG)-poly(dC)
counterpart. The different propagation scenarios are properly
illustrated using the time evolution of the first momentum of the
electronic occupation probability defined as
$\bar{n}(t)=\sum_n\,n\,|c_n(t)|^2$.  One can clearly observe  that
in the(dG)-(dC) case electron propagation proceeds unrestrictedly
and unidirectionally with uniform velocity (see Figure $4$).
Distinctly, the (dA)-(dT) electron moves itinerantly in a confined
region, comprising not more than five bases, around the starting
site restraining conductivity.
\begin{figure}
  \begin{center}
 \includegraphics[angle=-90,width=\singlefig]{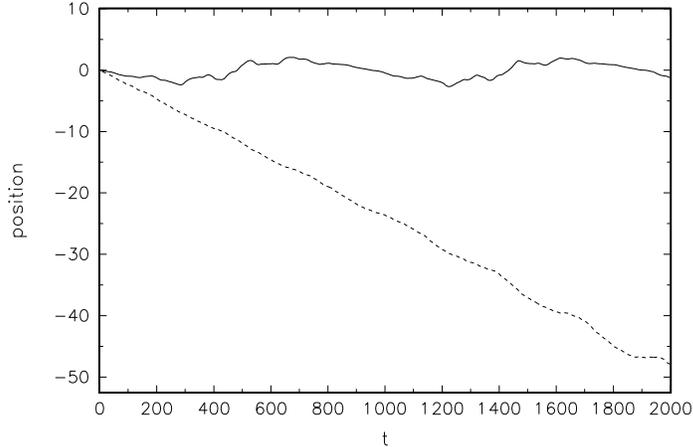}
  \end{center}
  \caption{Time evolution of the first momentum of
the electronic occupation probability. Full (dashed) line:
poly(dA)-poly(dT) (poly(dG)-poly(dC)) DNA polymer.}
\end{figure}

Regarding the energy storing capacity we monitored the normalized
participation number defined as
\begin{equation}
p(t)=\frac{P(t)}{P(0)}
\end{equation}
with
\begin{equation}
P(t)=\frac{1}{\sum_n\,\left|c_n(t)\right|^4} \,.
\end{equation}
Since the electronic wave function is normalized the electron
 breather is completely confined at a single site if
$p=1$ and is uniformly extended over the lattice if $p$ is of the
order $N$, viz. the number of lattice sites. Hence, $p$ measures
how many sites are excited to contribute to the electronic
breather pattern. From Figure $5$ we infer that the (dG)-(dC)
electron breather extension performs oscillations the maximal
amplitudes of which correspond to slight growth of the spatial
width up to merely $\simeq 1.2$ times its starting value. In
comparison, the width of the (dA)-(dT) electronic amplitude
pattern experiences stronger broadening leading to a more extended
electron state.
\begin{figure}
  \begin{center}
 \includegraphics[angle=-90,width=\singlefig]{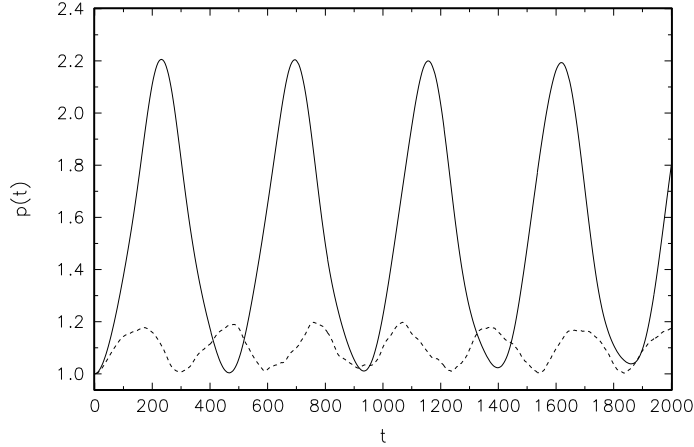}
  \end{center}
  \caption{The normalized participation number
$p(t)$. Full (dashed) line: poly(dA)-poly(dT) (poly(dG)-poly(dC))
DNA polymer.}
\end{figure}

Considering the energy exchange between the electronic and
vibrational subsystems we observe that this process evolves with
frequency $\Omega_r$. Hence, the dynamical distribution between
the electronic and vibrational energy is driven by the radial
harmonic vibrations. In particular, the amount of kinetic energy
flowing from the vibrational system into the electronic subsystem
(initially the electronic kinetic energy is zero) is governed by
the strength of the coupling $\alpha$. Apparently, in the
(dG)-(dC) case the energy sharing is by far more pronounced than
in the (dA)-(dT) case due to the fact that the coupling strengths
differ by almost an order of magnitude, that is
$\alpha_{(dG)-(dC)}=0.110$ and $\alpha_{(dA)-(dT)}=0.0154$. This
explains the large content of kinetic energy injected from the
vibrational degrees of freedom into the electronic ones inducing
high electron mobility. Furthermore, the $k$'s have opposite sign,
i.e. $k_{(dA)-(dT)}=-0.2591$ and $k_{(dG)-(dC)}=0.2234$, which
leads to the sign difference in the radial distortion patterns
$r_n^{(dA)-(dT)}$ and $r_n^{(dG)-(dC)}$.

An open question which is subject of further research is whether
the polarons survive at ambient temperature. In spite of the
values of the radial variables being small the polaron is a
compound object and  the fact that the transfer integral elements
$V_{n\,n-1}\approx 0.1\,\mathrm{eV}$ are larger than
$k_B\,T\approx 0.025\,\mathrm{eV}$ at $T=300\,\mathrm{K}$,
suggests that they would survive.
Equally the effect of an electric field
during the whole movement of the polaron is currently investigated.

In conclusion, we have found that conductivity in synthetically
produced DNA molecules depends on the type of the single base pair
of which the polymer is built of. While a polaron-like mechanism,
relying on the nonlinear coupling between the electron amplitude
and radial vibrations of the base pairs, is responsible for
long-range and stable ET in (dG)-(dC) DNA polymers, the
conductivity is comparatively weaker in the case of (dA)-(dT) DNA
polymers. Especially when it comes to designing synthetic
molecular wires these findings might be of interest. In fact,
recent experiments suggest that ET through DNA molecules proceeds
by polaron hopping \cite{Kawai}. Furthermore, our results comply
with the findings  of these experiments which show also that
poly(dG)-poly(dC) DNA polymers forms a better conductor than their
poly(dA)-poly(dT) counterparts.

 \section*{Acknowledgments}
One of the authors (D.H.) acknowledges support by the Deutsche
Forschungsgemeinschaft via a Heisenberg fellowship (He 3049/1-1).
Three of the authors (D.H., J.F.R.A. and F.P.) would like to
express their gratitude to the support under the LOCNET EU network
HPRN-CT-1999-00163. E.B.S. wishes to express his gratitude to
Prof. L. Nilsson (CSB NOVUM, Karolinska Institute, Sweden) for his
keen interest and stimulating discussions on the theme of this
work.


\end{document}